\documentstyle[11pt]{article}
\def\Journal#1#2#3#4{{#1} {\bf #2}, #3 (#4)}
\def\NPA{{\em Nucl. Phys.} A}
\def\PLB{{\em Phys. Lett.}  B}

\fussy
\begin{document}

\begin{center}
{\bf CONFINEMENT AND ENTROPY IN HEAVY ION COLLISIONS}
\vskip 4truemm

{\small \rm P. L\'EVAI\footnote{The talk was presented by P. L\'evai
at the Workshop on Quantum Chromodynamics
(3-8 June 1996, AUP Paris, France) and it will be published in
the Proceedings of the Workshop.}, J. ZIM\'ANYI }
\vskip 3truemm

{\small \it Research Institute for Particle and Nuclear Physics \\
H-1525 Budapest, 114. POB. 49, Hungary}
\end{center}

\noindent
{\footnotesize
We introduce a kinetical description of the hadronization
of semi-deconfined quark matter produced in heavy ion collisions.
A minimal microscopic model is considered in which
the produced quarks and anti-quarks are redistributed
into hadrons according to the additive quark model.
Assuming thermalized quark matter and hadron matter
we determine the appropriate parameters to describe
the properties of both matter before and after hadronization.
We find an adiabatic hadronization. 
}
\vskip 5truemm

In heavy ion collisions the evolution of hadronic fireball
was successfully described by the hadrochemical model \cite{hchem}.
This model is based on elementary processes in which hadron-hadron
collisions lead to the production of other type of hadrons. The 
time-evolution of the hadronic system was described
by a set of coupled differential equations in which the
time-dependent gain and loss terms were determined by means of
microscopical cross-sections for the elementary processes.
On the basis of the success of the hadrochemistry the quarkochemistry
model \cite{qchem} was developed to describe the evolution of
quark matter produced at ultra-relativistic energies. The
constituents of the quark-gluon plasma became the ingredi\-ents of the
microscopical processes and the cross sections could be obtained in
various ways.

In recent relativistic heavy ion collisions (see Pb+Pb at CERN SPS)
we expect the partial deconfinement of hadronic matter and the appearance of
some sort of {\bf semi-deconfined quark matter}~\cite{ALCORS95}. 
This deconfined state
will not be a quark-gluon plasma~\cite{QGP} (QGP)
with massless gluons and quarks in full equilibrium, but rather a mixture of 
quarks and diquarks coming from the wounded nucleons, 
and of course some newly produced quark-antiquark pairs, diquarks
and anti-diquarks. We assume that this mixture 
is thermalized or very close to it. Further assumption is that
the diquarks and anti-diquarks are such objects 
which are existing just temporally and 
they are not real degrees of freedom. 
Thus we have the quarks and anti-quarks as 
basic colored components. Considering the gluons we assume
that they are dressing the quarks which have some effective mass.
To describe the hadronization
of such a semi-deconfined quark matter, a kinetical model could be very
appropriate which would be the framework for the underlying microscopical
hadronization processes.
 
In a recent paper~\cite{ALCORS96} 
we introduced a model in which the elementary
processes contain the constituents of the above mentioned semi-deconfined
quark matter as incoming particles and the constituents of the hadron matter
as outgoing particles. In this way the
elementary processes of the phase transition from quark phase to hadron
phase can be modeled. This model, which is essentially an
extension of the earlier ALCOR model~\cite{ALCOR},
 may be called {\bf transchemistry},
i.e. the chemical-like description of the phase transition. 
If the thermalization remains valid during this transition then we
can follow the entropy production in the system. But this transition 
definitely will not be a fully equilibrated one, because the
microscopical processes will determine the number of the different
species, thus the system will stay far from any chemical 
equilibrium.

\begin{table}[tbp]
\caption{Particle numbers, entropy and CM energy (in GeV)
predicted by the ALCOR model for the Pb+Pb collision at
160 GeV/nucl. bombarding energy.}
\vspace{0.4cm}
\begin{center}
\begin{tabular}{|c|c|c|c|}    \hline
{\bf Pb+Pb} &
 {\bf NUMBER} & {\bf ENTROPY} & {\bf ENERGY} \\
\hline
 $h^{-}$&  730.41  & --- & ---    \\
\hline
 $\pi^+$&  603.87 & 1931   & 487    \\
\hline
 $\pi^0$&  618.95 & 1959   & 497    \\
\hline
 $\pi^-$&  634.68 & 1995   & 510    \\
\hline
 $K^+$  &  \ 84.15 & \ 490  & 121    \\
\hline
 $K^0$  &  \ 84.15 & \ 490  & 121   \\
\hline
 ${\overline K}^0$& \ 41.65    &\ 271 & \ 60   \\
\hline
 $K^-$  &  \ 41.65  & \ 271  & \ 60   \\
\hline
 $K^0_{S}$&  \ 62.90  &  ---  & ---    \\
\hline
 $p^+$  &  170.90  & 1137 & 393    \\
\hline
 $n^0$  &  188.57  & 1247  & 438    \\
\hline
 $\Sigma^+$&  \ 12.86  & \ 118  & \ 34    \\
\hline
 $\Sigma^0$&  \ 13.63  & \ 124  & \ 36    \\
\hline
 $\Sigma^-$&  \ 14.43  & \ 130  & \ 38    \\
\hline
 $\Lambda^0$& \ 68.20  & \ 526  & 178     \\
\hline
 $\Xi^0$&  \ \ 8.82  & \ \ 85  & \ 25    \\
\hline
 $\Xi^-$&  \ \ 8.89 & \ \ 86   & \ 25    \\
\hline
 $\Omega^{-}$&  \ \ 1.48  &  \ \ 18   & \ \ 5 \\
\hline
 ${\overline p}^-$&  \ 25.07 & \ 206   &\ 55    \\
\hline
 ${\overline n}^0$&  \ 25.07 & \ 206   &\ 55    \\
\hline
 ${\overline \Sigma}^-$&  \ \ 4.18  & \ \ 43   &\ 11 \\
\hline
 ${\overline \Sigma}^0$&  \ \ 4.18  & \ \ 43   &\ 11  \\
\hline
 ${\overline \Sigma}^+$&  \ \ 4.18  & \ \ 43   &\ 11     \\
\hline
 ${\overline \Lambda}^0$&  \ 20.93  & \ 181    &\ 53     \\
\hline
 ${\overline \Xi}^0$&  \ \ 5.98  & \ \ 59   &\  17    \\
\hline
 ${\overline \Xi}^+$&  \ \ 5.98  & \ \ 59   &\  17    \\
\hline
 ${\overline \Omega}^{+}$& \ \ 2.20&\ \ 27   &\ \ 8 \\
\hline
 Total: & --- & 11745 \    & 3267 \  \\
\hline
\end{tabular}
\end{center}
\end{table}

Considering the microscopical hadronization mechanisms here we will assume
that the production of the hadrons can be connected to the additive
quark model, e.g. a meson, which consists of quark and anti-quark,
$M_{ij}\equiv (q_i {\overline q}_j)$, will be produced from quark
and anti-quark.
Furthermore we can define
a minimal transchemistry model, in which we assume that all constituent
quarks and antiquarks are present in the semi-deconfined quark matter
from the beginning of the hadronization
and they will be confined into the hadrons in a minimal way, namely
$q_i + {\overline q}_j \Longleftrightarrow M_{ij}$, or especially
$q + {\overline q} \Longleftrightarrow \pi$.
In this case one can introduce the following rate-equations for the
pion, quark and antiquark numbers~\cite{ALCORS96}:
\begin{eqnarray}
{d \over {dt}} N_\pi (t) &=&
{ {D^{(\pi)} \cdot \langle \sigma^\pi_{q{\overline q}} v \rangle_{(t)} }
\over {V(t)} }
\cdot N_q (t) \cdot N_{\overline q} (t) - N_\pi(t) \cdot \Gamma_\pi \\
{d \over {dt}} N_q (t) &=&
- ~ { {D^{(\pi)} \cdot \langle \sigma^\pi_{q{\overline q}} v \rangle_{(t)} }
\over {V(t)} }
\cdot N_q (t) \cdot N_{\overline q} (t) + N_\pi(t) \cdot \Gamma_\pi \\
{d \over {dt}} N_{\overline q} (t) &=&
- ~ { {D^{(\pi)} \cdot \langle \sigma^\pi_{q{\overline q}} v \rangle_{(t)} }
\over {V(t)} }
\cdot N_q (t) \cdot N_{\overline q} (t) + N_\pi(t) \cdot \Gamma_\pi
\end{eqnarray}
Here $N_i(t)$ denotes the number of particle $i$, $V(t)$ is the
reaction volume, and
$\Gamma_\pi$ denotes the loss term for pions related to their
`decay' (or re-entering the semi-deconfined quark matter).
The $\langle \sigma^\pi_{q{\overline q}} v \rangle_{(t)}$ denotes
the momentum averaged elementary
hadronization rate for one hadronic degree of freedom
and $D^{(\pi)}$ is the spin degeneracy factor for the pion.
The eqs.(1)-(2)-(3) display that $N_\pi+ N_q=~const.$ and
$N_\pi+ N_{\overline q}=~const.$, which is just the number conservation 
for quarks and antiquarks
obtained from the additive quark model and the specific, unambiguous 
hadronization process $ q + {\overline q} \Longleftrightarrow \pi$.
The conservation of quark and antiquark number plays a crucial role
in the ALCOR model~\cite{ALCOR}, which is the linearized version
of the transchemistry. 

Note, that we can extend this minimal transchemistry model, introducing
different microscopical hadronization mechanisms, which will yield 
different coefficients in the above equations. 

Now we will apply the minimal transchemistry model and its linearized
version, the ALCOR, for the Pb+Pb collision at 160 GeV/nucl. 
bombarding energy. In Table~1 we display the produced 
particle numbers,  entropy and energy
for the outcoming hadron gas. 
Investigating the experimental results~\cite{Pb},
especially the transverse momentum distributions, one can find a
cylindrically symmetric thermalized hadronic fireball
to fit the measured particle distributions and numbers
(for more details see Ref.~\cite{ALCORS96}).
We obtained that the fireball can be characterized by
a central temperature $T_0=165 \ MeV$ and radial temperature
profile $T(r) = T_0 \cdot [1-(r/R)^2]$, a linear transverse
velocity profile with average transverse flow 
$<v_t>=0.38$, a transverse radius $R=14 \ fm$ and a
proper time $\tau=6 \ fm$.

The total entropy of the produced hadron gas is $S=11745$. Combining
this entropy value with the participant baryon number, $N_B=390$, one 
obtains that it is $S/N_B=30$ in the Pb+Pb collision. The total energy
is $E=3267 \ GeV$ (note that this energy contains both the thermal and
the flow energy) which is approximately equal to the expected one, 
$E_{CM}=390/2\cdot 17.4 = 3390 \ GeV$.

Now we assume that the initial quark matter 
was very much in the same condition
as the produced hadron matter, namely the central temperature
was $T= 165 \ MeV$ and there were 
similar temperature and velocity profiles which did not change during
hadronization. However, the hadronization takes some time, thus the
volume of the quark matter must have been smaller. 
We assume a hadronization time
$\Delta \tau = 4 \ fm$, thus the transverse radius 
was $R=10 \ fm$ at proper time $\tau= 2 \ fm$.
Furthermore, we need to reproduce the quark and antiquark numbers
obtained from ALCOR~\cite{ALCORS96} 
for this early, partially deconfined state:
$N_u = 1007$, $N_{\overline u}=464$,
$N_d=1089$, $N_{\overline d}=464$, $N_s=N_{\overline s}=236$.
We can reproduce these quark numbers at the above parameters, if
the quark effective masses are $m_q=150 \ MeV$ and $m_s = 300 \ MeV$.
(Note, that we are in the non-perturbative region, on the other hand
the expression $m_q= gT/\sqrt{6}$ yields the same effective mass
value for light quark at temperature $T=165 \ MeV$ and $\alpha_s = 0.4$.
For the effective strange quark mass 
one can use that $m_s \approx gT/\sqrt{6}
+ m_{s,0}$, where $m_{s,0}=150 \ MeV$ is the bare mass.)
For the semi-deconfined quark matter we obtained
total entropy $S=11390$ and  
total energy $E=3306\ GeV$, which are approximately the same values
than for the hadronic matter.

In conclusion 
we obtained that our minimal hadronization model, which based
on the idea of redistribution of quarks and anti-quarks, yields an adiabatic
hadronization process with $\Delta \tau = 4 \ fm$ hadronization time in the
Pb+Pb collision at SPS energy. 
A more extended numerical calculation 
containing other microscopical hadronization 
mechanisms is in progress to obtain
more details about the time-evolution of
hadronization and entropy production.
 
\section*{ Acknowledgment}
Discussion with B. M\"uller is acknowledged.
This work was supported by the National Scientific
Research Fund (Hungary), OTKA
No.T016206 and F019689. One of the author (P.L.) thanks for
the financial support obtained from the Soros Foundation and 
from the "For the Hungarian Science" Foundation of the Magyar
Hitelbank (MHB).


\begin{thebibliography}{99}
\parindent=8 truemm
\itemsep= -1mm
 
\bibitem{hchem} I.Montvay, J.Zim\'anyi,  \Journal{\NPA}{316}{490}{1979}.
 
\bibitem{qchem} T.S.Bir\'o, J.Zim\'anyi, \Journal{\PLB}{113}{6}{1982}.
 
\bibitem{ALCORS95} T.S. Bir\'o, P. L\'evai, J. Zim\'anyi,
Proceedings of the Strangeness'95 Conference, 1995 Tucson,
(AIP, Ed. J. Rafelski),
{\it AIP Conference Proceedings}  {\bf 340}, 405 (1995).

\bibitem{QGP} B. M\"uller, {\em The Physics of the Quark-Gluon Plasma,}
Lecture Notes in Physics, Vol. {\bf 225}, Springer, 1985.

\bibitem{ALCORS96} J. Zim\'anyi, T.S. Bir\'o, T. Cs\"org\H o, P. L\'evai, 
Proceedings of the Strangeness'96 Conference, 1996 Budapest,
{\em Heavy Ion Physics}  {\bf 4}, 15 (1996).

\bibitem{ALCOR} T.S. Bir\'o, P. L\'evai, J. Zim\'anyi, 
\Journal{\PLB}{347}{6}{1995}.
 
\bibitem{Pb} P. Seyboth,
 Proceedings of the XXV. International Symposium on Multiparticle
 Dynamics, Stara Lesna, 1995, World Scientific, Ed. by
 L. Sandor et al.

\end{thebibliography}
\end{document}